\documentclass[12pt]{article}
\usepackage[margin=1in]{geometry}
\usepackage{mathptmx,setspace,ragged2e,microtype,booktabs,tabularx,array,graphicx,amsmath,xurl}
\usepackage{longtable,fvextra}
\usepackage{float}
\usepackage[hidelinks]{hyperref}
\usepackage[numbers,square,sort&compress]{natbib}
\RaggedRight
\newcommand{\placement}[1]{\begin{center}\textit{[Insert #1 about here]}\end{center}}
\newcolumntype{Y}{>{\RaggedRight\arraybackslash}X}
\title{The AI Disclosure Paradox: Operational Disclosure and the Rise of Corporate AI Narratives}
\author{Abdullah Karasan\\
University of Maryland, Baltimore County (UMBC)\\
Leveragai, USA}
\date{November 2026}
\begin{document}
\maketitle

\begin{abstract}
AI-related corporate disclosure can expand while becoming less concentrated on current
use. Across U.S. SEC filings from 2017 to 2025, AI-related passage intensity
increased 6.3-fold between 2017--2022 and 2023--2025, whereas corrected
operational-disclosure intensity increased 3.3-fold. The operational share fell
from 25.76\% to 13.45\%; approximately 89\% of the net increase in passage
intensity fell outside the operational category. The decline persists within
sectors and filing types and in 10-K-only analysis. These estimates combine frozen
language-model predictions for 26,279 AI-related passages with an audit
of 1,000 annotated passages. The findings distinguish growth in
organizational attention from growth in implementation-specific claims: more
operational disclosure can coexist with a substantially lower operational
share. Operational disclosure was thus diluted, not displaced, by
nonoperational disclosure. Reading corporate technology narratives therefore requires attention
to both disclosure intensity and the organizational state described.
\end{abstract}


\section{Introduction}

Corporate discussions and announcements of artificial intelligence (AI) grew rapidly during 2017–2025. This growing attention raises a fundamental question: To what extent did corporate AI narratives describe technology already in operational use? Amid growing attention to AI, corporate filings can describe different topics such as operations, plans, pilot projects, or business risks. These topics signal different stages and forms of organizational engagement with AI.
This study therefore asks whether firms report using AI as much as their
growing discussion of it suggests. It addresses four questions. Did
disclosure of operational AI use keep pace as corporate attention to AI
expanded? What did the additional disclosure consist of? Was operational
disclosure displaced or only diluted by it? And do the answers hold
across sectors and filing types and under different ways of measuring
disclosure? These questions matter because organizations can respond to a socially
expected practice symbolically, by adopting its language, or
substantively, by implementing it \citep{bromley2012smoke}, because
information systems research treats the discourse surrounding a new
technology as distinct from its adoption \citep{swanson2025revisited},
and because recent studies find that firms' statements about AI and
their AI investment diverge \citep{li2025washing,liu2026glitters}.

In this study, it is found that statements describing operational AI use
increased, but AI-related disclosure overall grew even faster, reducing
the proportion that described operational use. Between 2017--2022 and
2023--2025, AI-related passages per 100 filings increased 6.3-fold,
while those describing operational AI use increased 3.3-fold.
Operational-use statements therefore fell from 25.76\% to 13.45\% of
AI-related passages. Approximately 89\% of the net increase in passage
intensity concerned content other than current operational use, and it
was concentrated in the general-or-uncertain category, not in plans or pilots.
Operational disclosure was therefore diluted, not displaced.

The share of AI-related passages describing operational use also declined when the sector and filing-type composition was held constant, falling by 13.39 percentage points. A 10-K-only analysis gives a decline
of 11.29 points. The finding therefore extends beyond a shift toward different
sectors or a dependence on earnings-release documents. 

The evidence comes from 26,279 AI-related passages identified through keyword screening in SEC filings
associated with historical S\&P 500 membership. A passage is classified as operational AI-use disclosure when it identifies an organizational actor, an AI technology, an application, and explicitly states that the technology is currently used or has been implemented. I use an audit of 1,000 manually annotated passages to correct errors in the language model’s estimate of how many passages describe operational AI use. 

This study contributes to research on AI-related disclosures in three ways. First, it distinguishes two dynamics in the relationship between operational and nonoperational disclosure: displacement, in which both operational-disclosure intensity and its share decline, and dilution, in which operational-disclosure intensity rises while its share falls. Unlike research on decoupling and AI washing, which examines discrepancies between organizational claims and actual practices, this distinction concerns the composition of disclosure itself and can be examined without independent evidence of implementation. Second, the study provides empirical evidence of disclosure dilution after 2022. Operational-use disclosures became more frequent but were outpaced by nonoperational disclosures. Most of the increase outside the operational category was concentrated in general-or-uncertain content, and the decline in operational share persisted after accounting for sector and filing-type composition. Third, the study demonstrates a transparent approach to estimating operational AI disclosure by combining explicit classification criteria, language-model predictions, and an audited sample of annotations to correct classification errors and quantify sampling uncertainty. Together, these contributions clarify what growing AI-related disclosure reveals about firms' reported operational use.
\section{Literature Review}

Prior studies show that AI-related disclosures have grown, particularly
since the release of ChatGPT, and that their content differs across
firms and reporting settings \citep{anantharaman2026disclosure,jia2026genai}. These studies show that AI-related disclosures cover operational use, future plans, opportunities, and risks. Growth in these disclosures therefore does not necessarily imply growth in reported operational use. 

A central issue is therefore what counts as a report of operational use.
Prior research separates AI-related disclosures by how specific they
are, whether they describe use or intention, and how they are worded
\citep{basnet2025narratives,nishant2024signals,shen2025adoption}. These
distinctions matter because a detailed plan to use AI is not a report
that AI is in use, and confident wording about AI capabilities is not
one either. When the question concerns operational AI use, statements
of current use must therefore be separated from plans, pilots, and
general discussion.

Recent research makes this distinction more explicit. \citet{shen2025adoption} separate applied AI, planned AI, and general disclosures when constructing firm-level indicators. \citet{balabanov2026deployment} distinguishes deployment from intentions, risk, and governance content. Together with research documenting expanding AI-related risk disclosures, these studies show why changes in total disclosure volume require attention to content \citep{anantharaman2026disclosure,balabanov2026deployment,shen2025adoption}. Operational-use disclosures and other AI-related disclosures do not need to expand at the same rate. Consequently, the frequency of operational-use disclosures and their proportion of all AI-related disclosures provide complementary information.

Institutional theory offers one possible reason why the two do not need to expand together. When a practice becomes socially expected, organizations gain legitimacy by adopting its language, and such symbolic adoption can run ahead of substantive adoption \citep{meyer1977institutionalized,fiss2006symbolic,bromley2012smoke}. Information systems research describes a related process for new technologies. A community's shared idea of what a technology is for, its organizing vision, is developed through discourse and shapes whether and how firms adopt the technology \citep{swanson1997organizing,swanson2025revisited}, association with a fashionable technology brings firms reputational benefits \citep{wang2010chasing}, and the acquisition of a technology can run well ahead of its deployment, a difference known as the assimilation gap \citep{fichman1999illusory}. Corporate disclosures respond to pressures of this kind \citep{darcy2022disclosures}, and market reactions to AI announcements depend on both substantive and rhetorical signals \citep{nishant2024signals}. Recent studies of AI washing compare firms' statements about AI with their AI investment or hiring and find that the two diverge \citep{li2025washing,liu2026glitters}. Those studies compare what firms say with data on what they do, such as
AI hiring or investment, whereas this study examines the composition of
disclosure itself. These accounts concern motives that disclosure text alone cannot establish. A passage that does not report operational use is not necessarily symbolic. It can describe a material risk, an investment, or an evaluation under way. This study therefore defines its constructs by what it measures, operational and nonoperational disclosure. Nonoperational disclosure can grow relative to operational disclosure in two ways. Under displacement, nonoperational disclosure takes the place of reports of use, so both the frequency of operational disclosures and their share fall. Under dilution, operational disclosure grows in absolute intensity but declines as a share of AI-related disclosure because nonoperational disclosure expands more rapidly. Symbolic adoption that stands in for implementation, the case mainly examined in research on decoupling, would appear as displacement. The two possibilities cannot be told apart from disclosure volume or from operational share alone; both must be observed together.

This interpretation also draws on research showing that the meaning of textual indicators depends on context. Financial vocabulary can differ substantially from general-language vocabulary, and firms can adapt their disclosures to anticipated readers and analytical tools \citep{loughran2011liability,cao2023machine}. More broadly, converting text into research variables requires choices about classification, aggregation, and the concept being measured \citep{gentzkow2019text}. For AI-related disclosures, these choices determine whether an indicator represents overall disclosure volume, intended activity, or reported operational use. Comparisons across periods must therefore consider both disclosure content and the reporting contexts represented in the sample.

Automated classification introduces a further distinction between identifying individual disclosures and estimating their prevalence. Language-model classifications can support large-scale analysis, but prediction accuracy alone does not establish that aggregate estimates or their confidence intervals are reliable \citep{egami2023surrogates,angelopoulos2023ppi}. Research on statistical inference using predicted data shows how sampled annotations can be combined with predictions to account for classification error. This literature provides a basis for estimating the prevalence of operational-use disclosures while recognizing uncertainty in the classification process.

Prior work thus establishes that AI-related disclosures differ in
content, but not whether reports of operational use were displaced or
diluted as those disclosures expanded. This study tests which occurred.
It identifies reports of operational use with explicit criteria applied
by a language model, corrects the resulting estimates with an audited
sample of annotations, and compares the two periods within sectors and
filing types.

\section{Data and Methodology}

In this study, U.S. Securities and Exchange Commission (SEC) filings are used for the firms
that belonged to the S\&P 500 at any point between January 2017 and
December 2025. The historical list contains 684 firms, of
which 653 had at least one filing meeting the study's selection criteria.
33,077 eligible filings were identified, and 33,064 of them were successfully processed. The processed dataset contains 4,175 annual reports
(Form 10-K), 12,648 quarterly reports (Form 10-Q), and 16,241 eligible
current reports (Form 8-K), including earnings releases and related
exhibits. Each filing is assigned to a year using its filing date and linked
to its firm using the SEC's Central Index Key (CIK), a unique firm
identifier.

I extracted the filing text and divided it into approximately
22.9 million usable sentences. Each passage contains a target sentence
and, where available, the sentences immediately before and after it.
These surrounding sentences provide context for interpreting the target
sentence.

The keyword search identified 26,420 AI-related passages, of which 26,279 were retained for analysis. The retained passages were then examined in context to determine whether they described operational AI use.

I also
checked the search procedure by inserting test sentences containing
listed AI terms and checking whether they were retrieved. These checks
assessed retrieval of the specified terms. They did not establish how
much AI-related content the keyword list might miss.

All successfully processed eligible filings enter the filing denominators,
including those without retained AI-related passages. Retained passages
occur in 1,630 annual reports, 1,986 quarterly reports, and 499 current
reports. A filing without retained passages is treated as having no
retrieved AI-related passages, rather than as evidence that the firm does
not use AI.

Each firm is assigned the same sector throughout the study period. 
To check whether the small number of AI-related passages in 8-K filings affects the results, the analysis is repeated using only 10-K annual reports.

Then, a passage is classified as describing operational AI use if it identifies an organizational actor, an AI technology, an application, and explicitly describes current use or completed implementation. The organizational actor must be the filing
firm or one of its units. For example, a disclosure that a firm uses
machine learning to route customer requests qualifies. A disclosure that
it plans to develop such a system does not. 

The classification procedure assigns one of five statuses to each passage. These are:
general or irrelevant, planned, pilot or trial, operational, or uncertain.
Only the operational category receives a value of 1 in the binary measure,
the other categories receive 0. The model also records functions, intended
outcomes, and provider roles, but those classifications are not used for
the population estimates reported here.

Qwen2.5-7B-Instruct is used and run locally with four-bit quantization.
The coding instructions, prompt, and model settings were refined using
the annotated passages described below. The keyword list, prompt,
model version, quantization settings, output-processing rules, and final
classification rule were then fixed and applied to all passages.

The model classifies each passage three times using fixed random seeds,
a temperature of 0.25, and a top-$p$ setting of 0.9. An operational
classification requires at least two of the three runs to identify
operational use and provide overlapping supporting text from the passage.
Fields without a resolved classification are marked uncertain.
Agreement across runs is a decision rule, not a probability that the
classification is correct. This approach draws on research using language
models for text annotation while recognizing that their classifications
can contain errors
\citep{gilardi2023annotation,mellon2024open,ziems2024css}.

In what follows, the model classifications are assessed against a human
annotator's audit of 1,000 passages, drawn in two stages. First, the
model was run on a pilot of 2,000 passages, and 200 of these were
selected for annotation, with passages the model classified as
operational and passages it was likely to have missed deliberately
overrepresented. Second, 800 of the remaining 26,079 passages were
randomly sampled without replacement within 150 groups defined by model
classification, period, filing type, and sector. The initial 200
passages are treated as fully observed: their annotations enter the
estimates directly, and the 800 sampled passages represent the rest of
the population. Appendix~\ref{app:sampling} reports the sampling groups.

The annotations for all 1,000 audited passages also informed the
refinement of the coding instructions, prompt, and model settings. This
does not affect the estimated shares for the full population, periods,
sectors, or filing types, which equal weighted estimates from the
annotations alone. It does mean that the model's precision, recall, and
specificity are measured on passages that informed its development and
are likely to be optimistic, and that the annual estimates, which depend
on the model's classifications, should be read with more caution.

A human annotator classified the initial 200 passages without seeing the model predictions. A subsequent AI-assisted review changed 126 annotation values across 68 passages. Because each passage has several annotations, these counts do not show how many passages changed between operational and nonoperational classifications. The primary analysis uses the resulting classifications for these 200 passages and the corrected classifications for the additional 800. The available records do not indicate whether AI assisted with corrections to the additional 800 passages.

An alternative set of corrected classifications is available for 300 overlapping passages and was used for sensitivity analysis. Its original annotation file duplicated the primary annotations. The comparison therefore assesses sensitivity to different corrected versions, but does not measure agreement between independent annotators.

I estimate the proportion of extracted AI-related passages that describe
operational AI use. For passage $i$, let $Y_i=1$ if the reviewed annotation
identifies operational use and $Y_i=0$ otherwise. Let $\widehat{Y}_i$ be
the model's classification prediction and let $S_0$ contain
the initial 200 audited passages and $S_1$ the additional 800 passages.
For each passage in $S_1$, $\pi_i$ is its known probability of selection
from the remaining passage population. The complete population,
$\mathcal{U}$, contains $N=26{,}279$ passages.

I begin with the model's operational-passage count and adjust it using
the differences between the reviewed annotations and the model
classifications:
\begin{equation}
\label{eq:operational_prevalence}
\widehat{p}
=
\frac{1}{N}
\left[
\sum_{i\in\mathcal{U}}\widehat{Y}_i
+
\sum_{i\in S_0}(Y_i-\widehat{Y}_i)
+
\sum_{i\in S_1}\frac{Y_i-\widehat{Y}_i}{\pi_i}
\right].
\end{equation}

The first term ($\sum_{i\in\mathcal{U}}\widehat{Y}_i$) in
Equation~\ref{eq:operational_prevalence} is the model's
operational-passage count across all extracted passages.
The second term ($\sum_{i\in S_0}(Y_i-\widehat{Y}_i)$)
replaces the model classifications with the reviewed annotations
for the initial 200 audited passages.
The third term ($\sum_{i\in S_1}(Y_i-\widehat{Y}_i)/\pi_i$)
estimates the correction needed for the remaining passage population.
Each difference in the additional sample receives weight $1/\pi_i$,
so passages with lower selection probabilities represent more passages.
Thus, if the model labels a passage nonoperational, but the annotator labels it operational, the correction increases the estimated count. This calculation adjusts the model-based estimate for classification error. It has the form of a prediction-powered difference estimator \citep{angelopoulos2023ppi}.

For each year, period, sector, or filing type, Equation~\ref{eq:operational_prevalence} estimates the number of operational-use passages belonging to that group. Only passages from that group contribute to the calculation, and the original audit weights are retained. Dividing this estimated count by the group’s total number of extracted passages gives its estimated operational share.

For the overall estimate, the model classifications cancel out, so the
result equals the weighted estimate from the audited annotations alone.
The procedure is therefore best described as an audit-corrected estimate
with model-assisted sampling: the model determines how passages are
grouped for sampling, and the annotations determine the estimate. The
model's classifications affect the estimates only for groups, such as
years, that cut across the sampling groups.
The same weights are used to estimate the shares of passages annotated
as planned, pilot, and general or uncertain.

Standard errors account for stratified sampling without replacement. Approximate 95\%
confidence intervals equal the estimate plus or minus 1.96 standard errors. These intervals account for sampling uncertainty conditional on the
reviewed annotations. They do not account for errors in those annotations.
Appendix~\ref{app:inference} provides the variance calculations.

The audit is also used to assess the model's classifications. Three
measures are reported, each calculated with the audit weights: precision, recall, and specificity. A precision of at least 85\% was set in advance as the
condition for using the model's classifications without correction.

This analysis uses three measures, namely, attention intensity, operational share, and operational intensity. Attention intensity is the number of
AI-related passages per filing. Operational share is the estimated
proportion of those passages describing operational AI use.
Operational intensity is the estimated number of operational-use
passages per filing. Each measure is calculated for a group $d$,
which can represent a year, period, sector, or filing type.
Let $C_d$ be the number of retained AI-related passages, $F_d$ the
number of successfully processed eligible filings, and
$\widehat{p}_d$ the estimated operational share in that group.
Attention intensity is calculated as $A_d=C_d/F_d$, and operational
intensity as $O_d=A_d\widehat{p}_d$. Both intensity measures are
multiplied by 100 when reporting results, so they express the number
of passages per 100 filings. These measures distinguish displacement
from dilution. Displacement implies that operational share and
operational intensity both fall. Dilution implies that operational share
falls while operational intensity rises.

The main analysis compares the three measures between 2017--2022 and
2023--2025. Annual estimates describe changes from year to year. The
shares of planned, pilot, and general or uncertain passages are compared
between the two periods in the same way. The comparison is repeated for
each filing type. Holm adjustments reduce the risk of identifying
differences by chance and are applied within each set of related tests:
the three filing types, the annual trend tests, and the sector
comparisons. Sector changes and sector rankings are examined as
exploratory analyses. Three checks assess whether the conclusions depend
on measurement choices: the period comparison using the model's
classifications alone, a comparison that excludes terms specific to
generative AI from the keyword list, and a firm-level comparison. The
last two are reported in Appendix~\ref{app:checks}.

Two decompositions separate the observed changes into components.
Both compare 2017--2022 ($t=0$) with 2023--2025 ($t=1$). The first decomposition examines the change in operational share.
It separates changes within sectors from changes in the proportion
of AI-related passages coming from each sector.

For sector $s$ in period $t$, let $\widehat{p}_{st}$ be its operational
share and $w_{st}$ its weight. This weight is the number of AI-related
passages from that sector divided by the total number of AI-related
passages in that period. Bars denote averages across the two periods.
The decomposition is
\[
\widehat{p}_1-\widehat{p}_0
=
\sum_s \bar{w}_s
(\widehat{p}_{s1}-\widehat{p}_{s0})
+
\sum_s (w_{s1}-w_{s0})\bar{p}_s,
\]
where $\widehat{p}_0$ and $\widehat{p}_1$ are the overall operational
shares in the earlier and later periods, respectively
\citep{kitagawa1955components}.

The first term measures changes in operational share within sectors.
It uses the same average sector weights for both periods, so changes
in sector weights do not affect this component. The second term
measures the contribution of changing sector weights. It holds each
sector's operational share at its average value across the two periods.
Together, the two terms equal the overall change in operational share.

The calculation is also performed for groups defined by both sector
and filing type. Only groups containing retained AI-related passages
in both periods are included. These comparisons account for changes
in the mix of sectors and filing types, but do not isolate changes
within individual firms.

The second decomposition separates changes in operational intensity
into contributions from attention intensity and operational share.
Attention intensity, $A_t$, is the number of AI-related passages
per filing in period $t$. Operational share, $\widehat{p}_t$, is the
proportion describing operational AI use. Their product gives
operational intensity:
\[
O_t=A_t\widehat{p}_t.
\]
Operational intensity can therefore change because attention intensity
changes, operational share changes, or both change. Their contributions
are separated as follows:
\[
O_1-O_0
=
(A_1-A_0)\bar{p}
+
(\widehat{p}_1-\widehat{p}_0)\bar{A},
\]
where
\[
\bar{p}=\frac{\widehat{p}_1+\widehat{p}_0}{2}
\qquad\text{and}\qquad
\bar{A}=\frac{A_1+A_0}{2}.
\]

The first term measures the contribution of changing attention
intensity, with operational share held at its average value.
The second term measures the contribution of changing operational
share, with attention intensity held at its average value.
Together, these terms equal the total change in operational intensity.

\section{Results}

Before discussing the findings, it would be worthwhile to see the model's classifications in Table~\ref{tab:validation}. Across all 26,279 passages, the corrected operational share is 16.53\%
(95\% confidence interval of 14.03\%--19.03\%), whereas the model alone
classifies 11.60\% as operational with a precision, recall and specificity of 80.6\%, 56.6\%, and 97.3\%, respectively. The model alone therefore rarely labels a
nonoperational passage as operational, but it misses almost half of the
operational ones. The correction addresses both errors. Using the
audited annotations, it removes the passages the model wrongly
classifies as operational, an estimated 590, and adds the operational
passages the model misses, an estimated 1,886. The estimated number of
operational passages thus rises from 3,049 to 4,345, or from 11.60\% to
16.53\% of all passages (Appendix~\ref{app:confusion}). Because
precision is below the 85\% threshold set in advance, all results below
use the corrected estimates. The model alone gives an operational share
of 18.41\% in 2017--2022 and 9.33\% in 2023--2025, a fall of 9.08
percentage points, compared with the corrected fall of 12.31 points.
The correction therefore raises the level of the operational share in
both periods; under either measure the share falls by about half. Two
checks support the corrected share. Replacing the primary annotations
with the alternative annotations for 300 passages gives 15.73\%. Using
only the 800-passage probability sample gives 16.15\%, with a period
change of $-12.01$ percentage points.

\placement{Table 1}

The results show that AI-related disclosures grew faster than
disclosures describing operational AI use. Attention intensity,
the average number of AI-related passages per 100 filings,
increased from 29.31 in 2017--2022 to 185.38 in 2023--2025.
Operational intensity, the estimated number of passages
describing operational AI use divided by the same number of filings, increased from
7.55 to 24.94, as Table~\ref{tab:period} shows. Thus, attention intensity
increased 6.3-fold, compared with a 3.3-fold increase in operational
intensity.
Their different rates nevertheless reduced operational share from 25.76\% to
13.45\%, a fall of 12.31 percentage points.

Attention intensity increased by approximately 156 AI-related
passages per 100 filings. Thus, approximately 88.9\% of the net
increase was outside the operational category. Table~\ref{tab:period} shows what this increase consisted of. The share
of general or uncertain passages rose from 66.36\% to 80.58\% of
AI-related passages, an increase of 14.22 percentage points (95\%
interval of $+7.22$ to $+21.22$). The share describing planned use
changed little, from 7.04\% to 5.97\%, and the share describing pilots
went from 0.84\% to zero; neither change can be distinguished from zero.
The growth outside the operational category was therefore concentrated
in the combined general-or-uncertain category, not in plans or pilots.

\begin{figure}[H]
\centering
\includegraphics[width=\textwidth,height=0.68\textheight,keepaspectratio]{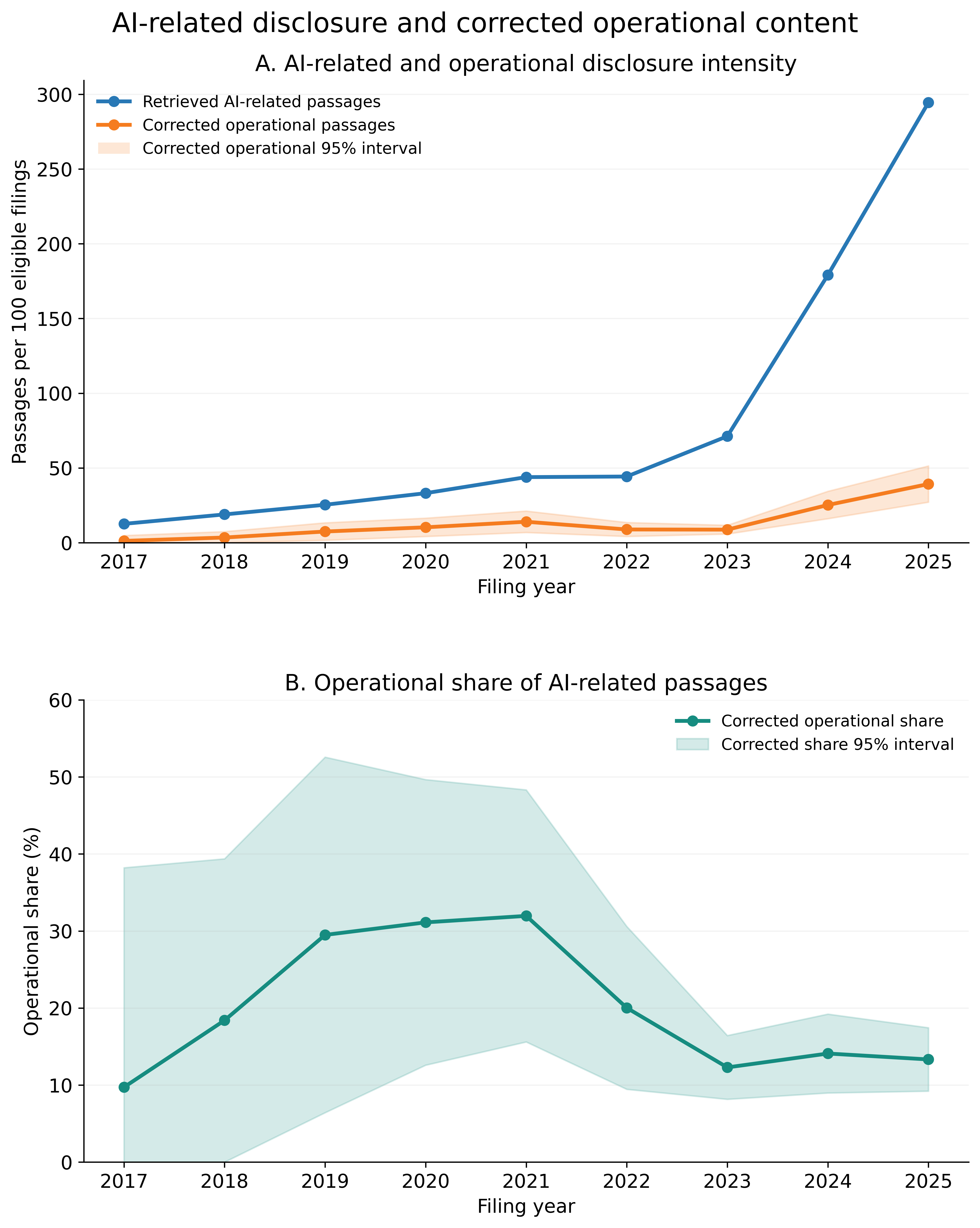}
\caption{AI-Related Disclosure Expanded Faster Than Operational Disclosure}
\label{fig:intensity}
\small\singlespacing
\textit{Note:} Intensities are retained passages per 100 processed eligible
filings. Shading and error bars use existing approximate 95\% confidence intervals.
The lower panel displays conditional operational share; early-year bounds
can cross zero. Annual points do not establish a steady linear decline.
\end{figure}

These figures raise a further question: how could operational intensity
rise while operational share fell? Table~\ref{tab:decomp} answers it by
separating the increase in operational intensity into the contributions
of attention intensity and operational share. Higher attention intensity
alone would have added 30.60 operational-use passages per 100 filings.
The decline in operational share removed 13.21 of them, leaving a net
increase of 17.39. More than two-fifths of the potential increase was therefore lost. Figure~\ref{fig:intensity}
shows both movements. In Panel A, the line for all AI-related passages
climbs steeply after 2022, whereas the line for operational passages
rises only modestly. In Panel B, the operational share drops from about
30\% in 2019--2021 to about 13\% in 2023--2025. Operational intensity
rose while operational share fell, which is the pattern of dilution, not
displacement.

The two series also moved at different times. Panel A of
Figure~\ref{fig:intensity} shows that AI-related passages rose from
44.29 to 71.24 per 100 filings in 2023, whereas corrected operational
passages remained nearly unchanged, moving from 8.87 to 8.76.
Operational passages began to rise in 2024, reaching 25.25 per 100
filings and then 39.26 in 2025. Operational disclosure intensity
therefore rose more sharply in 2024 than in 2023, one year after the
rise in AI-related disclosure. This is a descriptive annual pattern, not
a tested lag.

\placement{Table 2}
\placement{Figure 1}

\begin{figure}[!ht]
\centering
\includegraphics[width=\textwidth]{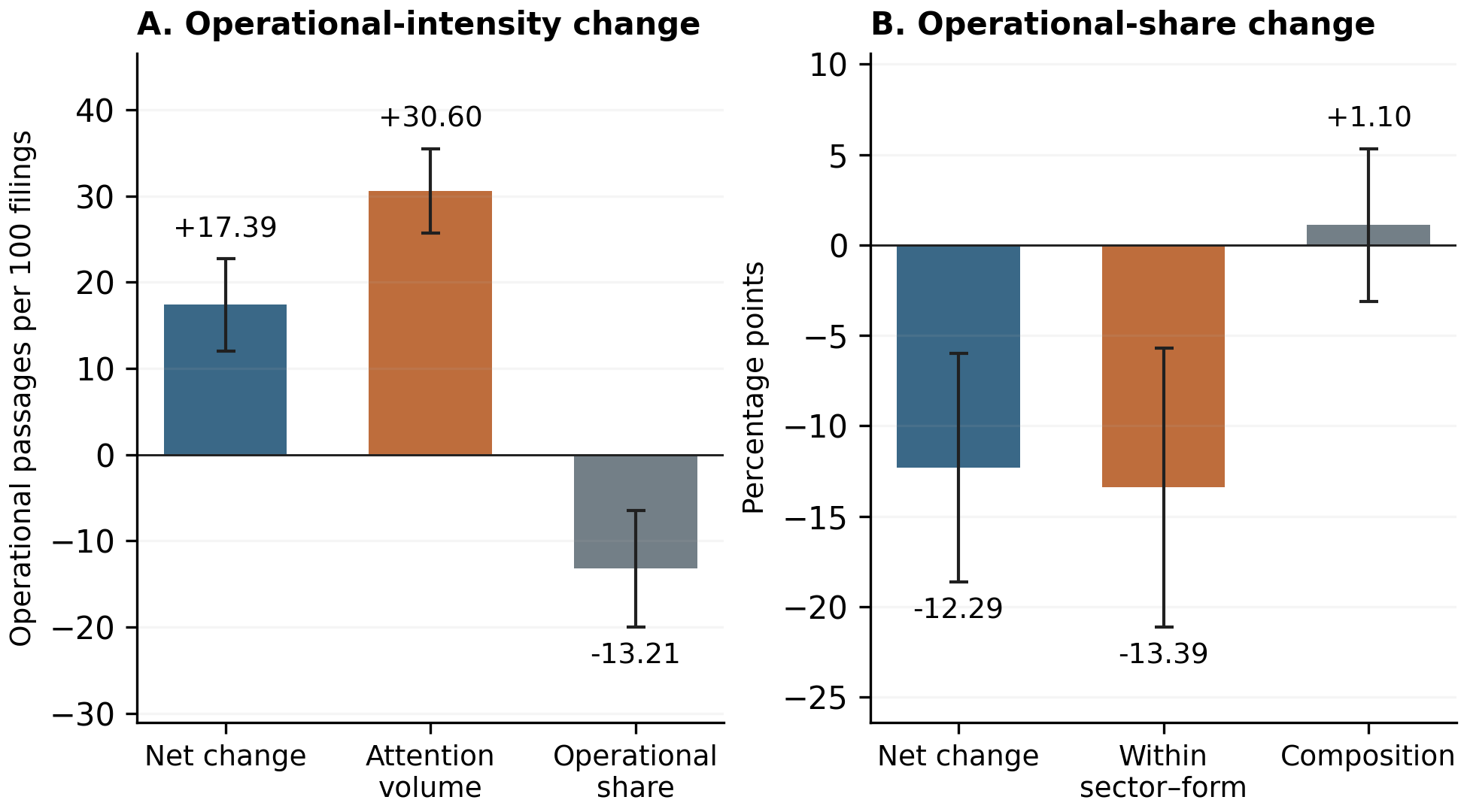}
\caption{Two Decompositions of the Disclosure Change}
\label{fig:decomp}
\small\singlespacing
\textit{Note:} The left panel decomposes operational intensity into attention-volume and operational-share components. The right decomposes operational
share within sectors and filing types that have AI-related passages in both periods. The panels have different units.
Bars and intervals reproduce existing estimates; neither panel is a causal
allocation.
\end{figure}

AI-related passages increased sharply, but the share describing
operational use fell. Panel B of Table~\ref{tab:decomp} shows that this
fall occurred within the same sector and filing type, where operational
share dropped by 13.39 percentage points (95\% interval of $-21.10$ to
$-5.68$). The change in where the passages come from had almost no
effect: $+1.10$ points ($-3.13$ to $+5.33$). The right panel of
Figure~\ref{fig:decomp} displays the same split.

\placement{Table 3}
\placement{Figure 2}

Table~\ref{tab:forms} reports the results separately for each filing
type. In 10-K annual reports, AI-related passages rose from 119.54 to
842.80 per 100 filings, about sevenfold, while operational passages
rose from 31.34 to 125.83, about fourfold. Operational share in 10-K
filings therefore fell from 26.22\% to 14.93\%, a drop of 11.29
percentage points (95\% interval: $-18.39$ to $-4.18$; Holm-adjusted
$p=.006$). The 10-Q quarterly reports show a similar drop of 14.15
points ($-27.11$ to $-1.18$). The fall in operational share is
therefore present in annual and quarterly reports and does not depend
on 8-K filings. The 8-K estimate for 2023--2025 rests on only 66
passages and is too imprecise to interpret.

\placement{Table 4}

The operational share settled at a lower level after 2022; it did not
keep falling every year. Panel B of Figure~\ref{fig:intensity} shows
the annual shares, which stay between 12.29\% and 14.10\% from 2023 to
2025, even as AI-related passages keep rising in Panel A. The annual
trend across 2017--2025 is not statistically significant (adjusted
$p=.524$).

Sectors do not differ detectably in how much their operational share
changed. A joint test of equal changes across sectors is not
significant ($\chi^2(10)=7.61$, $p=.667$). Three pairs of sectors
change rank order between the two periods, but none of these changes
is significant after Holm adjustment. The findings therefore concern
the overall fall in operational share, not differences among sectors.

Two further checks are reported in Appendix~\ref{app:checks}. Removing the
terms specific to generative AI from the keyword list leaves the fall in
operational share almost unchanged, at 12.42 percentage points. At the
firm level, among firm-years with AI-related passages, the proportion
containing at least one passage that the model classifies as operational
fell from 50.51\% to 37.95\%. The number of such firm-years nevertheless
rose, from 398 over six years to 373 over three years, or from about 66
to about 124 per year. This is suggestive of dilution at the firm level
but does not establish it. The check uses the model's classifications
without correction, counts only firm-years with AI-related passages,
compares periods of different length, and does not follow the same firms
over time.

\section{Discussion}
The findings show why growth in AI-related disclosures must be interpreted
alongside their content. Operational intensity increased, but operational
share declined because attention intensity grew faster. Operational-use
disclosures therefore became more frequent per filing while accounting
for a smaller proportion of all AI-related disclosures.

The pattern is dilution rather than displacement: operational
disclosure grew in absolute intensity while declining as a share of
AI-related disclosure. Institutional theory offers one possible
explanation. After 2022, AI became a topic that firms were expected to
address, and the fall in operational share appeared within sectors and
filing types, without detectable differences among sectors, as would be
expected of a field-wide pressure \citep{meyer1977institutionalized}.
In information systems terms, the period resembles the rapid spread of
an organizing vision for AI \citep{swanson1997organizing}, and the
difference between discussion and reports of use is the disclosure
counterpart of an assimilation gap \citep{fichman1999illusory}.
Yet the additional disclosure did not take the place of reports of use,
which tripled per filing. Nor was it concentrated in commitments to
future use, because the shares of planned and pilot passages did not
rise.
Reports of use rose more sharply in 2024 than in 2023, a year after
discussion expanded, and from 2023 the operational share remained stable
at a lower level. Operational and nonoperational disclosure therefore
grew together, with the nonoperational growing faster. This differs from
the decoupling described in prior work, in which symbolic adoption stands
in for implementation \citep{fiss2006symbolic}. It also differs from AI
washing, in which statements about AI are not matched by investment
\citep{li2025washing,liu2026glitters}: here, reports of use increased
alongside the additional disclosure. The evidence does not establish
that legitimacy-seeking caused this pattern. Nonoperational passages
include risk statements, investment disclosures, and evaluations that
can be informative in their own right, and the estimates concern what
firms disclose, not whether their statements match their practices.

This interpretation builds on research distinguishing different types of
AI-related disclosures
\citep{basnet2025narratives,shen2025adoption,balabanov2026deployment}.
The study combines that distinction with estimates adjusted for
classification error and comparisons of attention intensity, operational
share, and operational intensity. This approach helps researchers assess
what growing AI-related disclosures reveal about reported operational use.
Interpreting growth in attention intensity as equivalent growth in
operational intensity would overstate the latter in this dataset.
Operational share alone, however, misses the increase in operational
intensity. For firms, descriptions of current applications and
implementation status provide information that disclosure volume alone
cannot convey.

The sector and filing-type comparisons strengthen this interpretation.
The decline in operational share persists when sector and filing-type
weights are held fixed and also appears in 10-K annual reports. It is
therefore not explained solely by changes in reporting composition or
reliance on 8-K filings. These comparisons do not, however, isolate changes
within individual firms or demonstrate that sectors changed differently.

The main limitations concern undisclosed use and repeated passages. The
estimates cover only what firms report in passages retrieved by the AI
keyword search, so AI use that firms do not disclose, or describe
without the listed terms, is not captured. In addition, several
passages can describe the same application, so the results do not
measure firm adoption rates or the number of distinct deployments. The
comparisons are also limited by fixed sector classifications and
incomplete verification of 8-K retrieval. The estimates for planned,
pilot, and general passages rest on fewer audited passages than the
operational estimate and are less precise. Dilution is established here
for aggregate disclosure; at the firm level the evidence is only
suggestive, and a within-firm analysis of a balanced panel would be
needed to establish it. Finally, the audited
annotations also informed the development of the classification
procedure, so the model's reported precision, recall, and specificity
are likely to be optimistic.

\section{Conclusion}

Firms disclose far more about AI than they did before 2023, but most of
that additional writing does not describe AI in use. Between 2017--2022
and 2023--2025, AI-related passages in SEC filings grew about sixfold,
while passages describing operational use grew about threefold. As a
result, the portion of AI-related disclosure that reports operational use
fell by half, from roughly one passage in four to about one in seven.

The volume of AI disclosure alone therefore does not show how much of it reports operational use.
Those who read more AI language as more AI use would
overstate the growth in reported use by a wide margin. The opposite
reading is also wrong. Reports of operational use did increase
substantially, but they were outpaced by AI-related passages that do
not report use. Disclosure volume and operational share must be read
together, because either one alone gives a distorted picture. The change
is also broad. It occurred within sectors and filing types, showed no
statistically detectable differences across sectors, and appears in
annual reports. The lower operational share persisted through 2025 and
did not keep falling. Nonoperational disclosure diluted reports of use;
it did not displace them.

The study also shows the value of pairing language-model classification
with an audited sample. Correcting the model's classifications with the
audit raised the estimated operational share by nearly five percentage
points, a difference large enough to change conclusions drawn from the
text. The same procedure can be applied to other text-based measures of
technology adoption, giving estimates with stated uncertainty.

For firms, the findings imply that specific descriptions of current
applications carry information that general statements about AI do not.
For readers of corporate filings, they imply that the useful question is
what a firm says it is doing with AI, not how often it mentions AI.
These conclusions concern what firms disclose; they do not verify that
the disclosed systems are deployed or that they deliver benefits.

\section*{Use of Artificial Intelligence}

Qwen2.5-7B-Instruct, an open-weight language model, was used to classify
passages, as described in the Methodology, because the corpus is too
large to classify by hand. Claude was used to assist with manuscript
editing and language improvement, with the review of analysis code, and
with the review of passage annotations described in the Methodology, in
order to improve clarity and to check the work. The author reviewed and
revised all content and takes full responsibility for it.

\clearpage
\bibliographystyle{plainnat}
\bibliography{references}

\clearpage
\appendix
\section{Inference}
\label{app:inference}

For domain $d$, define $r_{id}=I(i\in d)(Y_i-\widehat Y_i)$. Under independent
simple random sampling without replacement within sampling groups, the
covariance estimator is
\[
\widehat{\operatorname{Cov}}(\widehat p_d,\widehat p_e)
=\frac{1}{N_dN_e}\sum_{h}
 \frac{N_h^2(1-n_h/N_h)}{n_h}s_h(r_d,r_e),
\]
where $N_h$ and $n_h$ are the population and audit sizes of group $h$, and
$s_h$ is the sample covariance with denominator $n_h-1$. The covariance
applies to the 800 sampled passages, the initial 200 passages are treated
as fixed. The interval for the difference between the corrected and
model-alone shares likewise treats the model's classifications as fixed.

For domains nested in the sampling cells, the model terms cancel. Overall,
\[
\widehat p=\frac{1}{N}\left[\sum_{i\in S_0}Y_i+
\sum_{i\in S_1}\frac{Y_i}{\pi_i}\right].
\]
This identity concerns the same weighted audit, not the benefit of predictions
for audit allocation. Annual domains need not be nested.

\setcounter{table}{0}
\renewcommand{\thetable}{A\arabic{table}}
\renewcommand{\theHtable}{A\arabic{table}}

\section{Stable-lexicon and firm-level checks}
\label{app:checks}

\subsection{Stable-lexicon check}

The benchmark candidate frame used one lexicon frozen before classification
and applied without period-specific changes. As an additional robustness
check, the stable-lexicon specification removes terms specific to the
post-ChatGPT generative-AI vocabulary: \textit{generative AI},
\textit{generative artificial intelligence}, \textit{gen AI}, \textit{large
language model(s)}, \textit{LLM(s)}, \textit{foundation model(s)}, and
\textit{transformer model(s)}. A passage is retained if it contains at least
one other frozen term. Human correction is re-estimated for this restricted
domain with the original probability-audit weights, it is not a model-only
subsample calculation.

\begin{table}[htbp]
\centering\small\singlespacing
\caption{Stable-lexicon robustness by period}
\begin{tabularx}{\textwidth}{@{}Xrrr@{}}
\toprule
Quantity & 2017--2022 & 2023--2025 & Post minus pre \\
\midrule
Stable-lexicon candidate passages & 6,568 & 18,703 & --- \\
Attention per 100 eligible filings & 29.27 & 175.98 & --- \\
Corrected operational passages per 100 filings
  & 7.55 & 23.53 & --- \\
Corrected operational share
  & 25.79\% & 13.37\% & $-12.42$ pp \\
Approximate 95\% interval for share & [20.14, 31.45] &
  [10.55, 16.19] & [$-18.75$, $-6.10$] pp \\
\bottomrule
\end{tabularx}
\end{table}

The substantive result is unchanged. The benchmark decline is $-12.31$
percentage points; under the stable lexicon it is $-12.42$ points. Removing
the emerging terms reduces post-period attention and corrected operational
intensity, but does not explain the decline in the operational share.
The check addresses vocabulary that entered use after 2022. It does not
test whether the lexicon misses earlier wording for similar technologies.

\subsection{Exploratory firm-year proxy}

The firm-year check aggregates frozen passage predictions, not human-corrected
firm-year adoption. The denominator is candidate-observed firm-years. The one-evidence proxy requires at least
one model-positive passage. The stricter proxy requires model-positive
passages in at least two distinct SEC accessions in the same firm-year.

\begin{table}[htbp]
\centering\small\singlespacing
\caption{Model-label firm-year disclosure proxies}
\begin{tabularx}{\textwidth}{@{}Xrrr@{}}
\toprule
Quantity & 2017--2022 & 2023--2025 & Change \\
\midrule
Candidate-observed firm-years & 788 & 983 & --- \\
At least one positive passage & 50.51\% (398) & 37.95\% (373)
  & $-12.56$ pp \\
Positive passages in at least two accessions & 14.47\% (114) &
  11.70\% (115) & $-2.77$ pp \\
\bottomrule
\end{tabularx}
\end{table}

These proxies describe disclosure persistence within the candidate frame.
They are not estimates of the percentage of firms adopting AI. Candidate-free
firm-years are outside the denominator, repeated text can represent one
deployment, and passage-level error correction does not transfer mechanically
to a nonlinear at-least-one firm-year indicator.

\section{Annotation Guidelines}
\label{app:codebook}

\subsection{Unit and binary target}

The annotator labeled each displayed target passage using only the
displayed context. The binary reference outcome equals one only when the
status is operational. Operational status requires all four
conditions: (1) the filing firm is the actor; (2) an AI technology is
identified; (3) an action and object are stated; and (4) current or completed
realization is established. Current AI operating inside the firm's product or
service counts, including AI work performed by employees to deliver a service.
A product name, launch, availability, sales statement, demand statement, or
development plan alone does not establish operation.

\subsection{Status codebook}

{\singlespacing\small
\begin{longtable}{@{}p{0.24\textwidth}p{0.70\textwidth}@{}}
\toprule
Label & Coding rule \\
\midrule
\endhead
\texttt{operational} & Current or completed AI operation in the filing
firm's own product, service, workflow, or decision. Present-tense operating
capabilities qualify when tied to the firm's current offering. \\
\texttt{pilot\_trial} & Proof of concept, limited test, pilot, beta, or
shadow deployment without evidence of broader current operation. \\
\texttt{planned} & Intention, evaluation, development plan, or future
commitment without established current use. \\
\texttt{irrelevant\_general} & Rhetoric, risk boilerplate, market discussion,
non-AI meanings, product demand without operating capability, or use by a
customer, supplier, competitor, or industry actor outside the filing firm's
offering. \\
\texttt{uncertain} & The displayed evidence does not permit a defensible
choice among the preceding categories. \\
\bottomrule
\end{longtable}
\addtocounter{table}{-1}
}

\section{Probability-audit sampling groups}
\label{app:sampling}

The original 200-item gold audit was drawn after the 2,000-item model pilot.
Its conditional allocation was enriched for model positives and likely
false-negative strata. The final augmentation excluded those 200 exact audited
units, then sampled 800 of the remaining 26,079 candidate passages by simple
random sampling without replacement within prediction-status $\times$ period
$\times$ filing-form $\times$ sector cells. There were 150 realized
augmentation cells. Inclusion probabilities and inverse weights were retained
at the actual-cell level.

\begin{table}[htbp]
\centering\small\singlespacing
\caption{Audit-stage group sizes}
\begin{tabularx}{\textwidth}{@{}Xrrl@{}}
\toprule
Stage and group & Frame $N$ & Audit $n$ & Design note \\
\midrule
Original pilot: model positive & 223 & 120 &
Conditional draw within the 2,000-item pilot \\
Original pilot: negative boundary & 37 & 21 &
At least one positive model vote \\
Original pilot: negative zero-vote & 1,740 & 59 &
No positive model vote \\
\addlinespace
Augmentation: model positive & 2,929 & 250 &
Allocated across period/form/sector cells \\
Augmentation: negative boundary & 447 & 200 &
Allocated across period/form/sector cells \\
Augmentation: negative zero-vote & 22,703 & 350 &
Allocated across period/form/sector cells \\
\midrule
Final audited total & 26,279 & 1,000 &
200 exact audited units plus 800 probability-audited remainder units \\
\bottomrule
\end{tabularx}
\end{table}

The table reports marginal group sizes for transparency. Analysis used
$\pi_i=n_h/N_h$ and $w_i=1/\pi_i$ from each actual augmentation cell $h$, not
the marginal ratios in the table. The 200 original audited units had analysis
weight one after exclusion from the augmentation frame; their initial
two-stage selection probabilities describe how that development/gold audit
was formed, not their final remainder-cell weight.

Annotations for all 1,000 audited passages also informed the refinement of
the coding instructions, prompt, and model settings. The estimated shares for
the full population, periods, sectors, and filing forms equal weighted
estimates from the annotations alone and are unaffected. The performance
measures in the next section are computed on passages that informed
development and are likely to be optimistic.

\section{Sampling-weighted confusion matrix}
\label{app:confusion}

\begin{table}[htbp]
\centering\small\singlespacing
\caption{Model prediction versus human binary reference label}
\begin{tabular}{@{}lrrr@{}}
\toprule
Human reference & Model operational & Model nonoperational & Weighted total \\
\midrule
Operational & 2,458.69 & 1,885.85 & 4,344.55 \\
Nonoperational & 590.31 & 21,344.15 & 21,934.45 \\
\midrule
Weighted total & 3,049.00 & 23,230.00 & 26,279.00 \\
\bottomrule
\end{tabular}
\end{table}

Weighted performance implied by this matrix is 80.64\% precision, 56.59\%
recall, 97.31\% specificity, and 90.58\% accuracy. Values are population-weighted
estimated counts and need not be integers.

\section{Reproducibility files}

For the sake of reproducibility, all the codes are available in github repository here\footnote{\url{https://github.com/abdullahkarasan/operational-ai-disclosure-sec}.}

\setcounter{table}{0}
\renewcommand{\thetable}{\arabic{table}}
\renewcommand{\theHtable}{\arabic{table}}

\clearpage
\section*{Tables}

\begin{table}[!ht]
\centering\small\singlespacing
\caption{Model Classifications Compared with the Audit}
\label{tab:validation}
\begin{tabularx}{\textwidth}{@{}p{.46\textwidth}p{.19\textwidth}Y@{}}
\toprule
Quantity & Estimate & Interpretation \\
\midrule
Operational share, model alone & 11.60\% & Model classifications without correction \\
Operational share, corrected & 16.53\% & Primary annotations \\
95\% confidence interval & 14.03\%--19.03\% & Sampling uncertainty \\
Corrected minus model alone & $+4.93$ pp & 95\% confidence interval [$+2.43$, $+7.43$] \\
Period change, model alone & $-9.08$ pp & Corrected change: $-12.31$ pp \\
Precision & 80.6\% & Below the 85\% threshold set in advance \\
Recall & 56.6\% & Operational passages the model identifies \\
Specificity & 97.3\% & Nonoperational passages the model identifies \\
Corrected share, alternative annotations & 15.73\% & Substitution on 300 passages; not independent agreement \\
Corrected share, 800-passage sample only & 16.15\% & Excludes the initial 200 passages \\
\bottomrule
\end{tabularx}
\end{table}

\clearpage
\begin{table}[!ht]
\centering\small\singlespacing
\caption{More Operational Disclosure, a Lower Operational Share}
\label{tab:period}
\begin{tabularx}{\textwidth}{@{}p{.35\textwidth}p{.16\textwidth}p{.16\textwidth}Y@{}}
\toprule
Quantity & 2017--2022 & 2023--2025 & Comparison \\
\midrule
Processed filings & 22,436 & 10,628 & Denominator of the rates \\
AI-related passages & 6,577 & 19,702 & Retained by the keyword search \\
\addlinespace
AI-related passages per 100 filings & 29.31 & 185.38 & $6.32\times$ \\
Operational passages per 100 filings & 7.55 & 24.94 & $3.30\times$ \\
Passages outside operational category per 100 filings & 21.76 & 160.44 & $+138.68$ \\
Operational share & 25.76\% & 13.45\% & $-12.31$ pp [$-18.59$, $-6.02$] \\
Planned share & 7.04\% & 5.97\% & $-1.07$ pp [$-6.22$, $+4.08$] \\
Pilot share & 0.84\% & 0.00\% & $-0.84$ pp [$-3.75$, $+2.06$] \\
General or uncertain share & 66.36\% & 80.58\% & $+14.22$ pp [$+7.22$, $+21.22$] \\
\bottomrule
\end{tabularx}
\par\smallskip\footnotesize
\textit{Note:} Brackets are approximate 95\% confidence intervals.
\end{table}

\begin{table}[!ht]
\centering\small\singlespacing
\caption{Two Decompositions of the Change Between 2017--2022 and 2023--2025}
\label{tab:decomp}
\begin{tabularx}{\textwidth}{@{}p{.48\textwidth}p{.17\textwidth}Y@{}}
\toprule
Component & Estimate & Approximate 95\% confidence interval \\
\midrule
\multicolumn{3}{@{}l}{\textit{A. Operational passages per 100 filings}} \\
Net change in operational intensity & $+17.39$ & [$+12.03$, $+22.75$] \\
Contribution of attention intensity & $+30.60$ & [$+25.69$, $+35.50$] \\
Contribution of operational share & $-13.21$ & [$-19.96$, $-6.46$] \\
\addlinespace
\multicolumn{3}{@{}l}{\textit{B. Operational share, percentage points}} \\
Total change & $-12.29$ & [$-18.61$, $-5.97$] \\
Change within the same sector and filing type & $-13.39$ & [$-21.10$, $-5.68$] \\
Change in where the passages come from & $+1.10$ & [$-3.13$, $+5.33$] \\
\bottomrule
\end{tabularx}
\par\smallskip\footnotesize
\textit{Note:} The two panels are separate calculations with different units.
Panel B uses sectors and filing types that have AI-related passages in both
periods; these cover 98.56\% of 2017--2022 passages and 99.75\% of 2023--2025
passages. Across all passages, the change in operational share is $-12.31$
points.
\end{table}

\clearpage
\begin{table}[!ht]
\centering\small\singlespacing
\caption{Coverage and Change in Operational Share by Filing Type}
\label{tab:forms}
\begin{tabularx}{\textwidth}{@{}p{.38\textwidth}YYY@{}}
\toprule
Quantity & 10-K & 10-Q & 8-K \\
\midrule
Processed filings & 4,175 & 12,648 & 16,241 \\
Filings with AI-related passages & 1,630 & 1,986 & 499 \\
AI-related passages & 14,639 & 10,672 & 968 \\
Operational share, 2017--2022 & 26.22\% & 25.70\% & 24.19\% \\
Operational share, 2023--2025 & 14.93\% & 11.56\% & 3.03\% \\
Change, percentage points & $-11.29$ & $-14.15$ & $-21.16$ \\
Approximate 95\% confidence interval & [$-18.39$, $-4.18$] & [$-27.11$, $-1.18$] & [$-71.13$, $+28.82$] \\
\bottomrule
\end{tabularx}
\par\smallskip\footnotesize
\textit{Note:} The 10-K change has unadjusted $p=.002$ and Holm-adjusted
$p=.006$ across the three filing types. The 2023--2025 estimate for 8-K
filings uses 66 passages. A filing without AI-related passages is not
evidence that the firm does not use AI.
\end{table}

\clearpage
\section*{Figures}

\clearpage
\end{document}